# Coronavirus (COVID-19): ARIMA based time-series analysis to forecast near future

**Hiteshi Tandon[1]\*, Prabhat Ranjan[2], Tanmoy Chakraborty[3]\*, Vandana Suhag[4]**

[1]Department of Chemistry, Manipal University Jaipur, Jaipur 303007, India

[2]Department of Mechatronics Engineering, Manipal University Jaipur, Jaipur 303007, India

[3]Department of Chemistry, School of Engineering, Presidency University, Bengaluru 560064, India

[4]Department of Applied Sciences, BML Munjal University, Gurugram 122413, India

\*Corresponding authors. Email: tanmoychem@gmail.com, tanmoychakraborty@presidencyuniversity.in (T. Chakraborty); hiteshitandon@yahoo.co.in (H. Tandon)

**ABSTRACT:** COVID-19, a novel coronavirus, is currently a major worldwide threat. It has infected more than a million people globally leading to hundred-thousands of deaths. In such grave circumstances, it is very important to predict the future infected cases to support prevention of the disease and aid in the healthcare service preparation. Following that notion, we have developed a model and then employed it for forecasting future COVID-19 cases in India. The study indicates an ascending trend for the cases in the coming days. A time series analysis also presents an exponential increase in the number of cases. It is supposed that the present prediction models will assist the government and medical personnel to be prepared for the upcoming conditions and have more readiness in healthcare systems.

## 1. INTRODUCTION

The pandemic of 2019-nCov commenced from December 2019 in Wuhan, China and has caused extreme havoc in almost the whole world.[1,2] 2019-nCoV or COVID-19, commonly known as Coronavirus, is a novel highly contagious virus belonging to *Coronaviridae* family that has been suspected to be transmitted to humans from animals. This virus causes mild to severe respiratory illness and death.[3] This pandemic has engulfed 185 countries/regions in merely four months infecting 1,949,210 people and taking the death toll to 123,348.[4,5] However, the premature cases show the infection is less severe as compared to other coronaviruses such as SARS-CoV (Severe Acute Respiratory Syndrome Corona Virus) and MERS-CoV (Middle East Respiratory Syndrome Corona Virus), the cases of rapid human-to-human transmission signify that 2019-nCoV is highly infectious than others.[6] Although a local seafood market in Wuhan is believed to be the source of exposure,[7] the scope of occurrence of this disease is not clear since its occurrence at present is so dynamic.[3] An apparent variation is present in epidemiological examinations and detection abilities performed by different countries for detecting infected cases.[8] Presently, the highest cases of 2019-nCoV infections have been reported in US, however, the cases are abruptly rising in Spain, Italy, France and Germany daily.[4] China, the place of origin of the disease, is now receiving a very few cases.[4] The first case of coronavirus infection in India was reported on 30 January 2020 in Kerela, which was an imported case from Wuhan city of China.[9] In the initial phase the spread was extremely slow and only 3 people were positive for more than a month. However, the numbers started rising exponentially after one month and continue to do so. The numbers in India have reached up to 10,453 for confirmed COVID-19 infected cases with 358 deaths and 1181 recoveries as reported on 13 April 2020.[4] At present, there is neither a treatment nor a vaccination for the COVID-19 infection. Currently, it is a major health crisis around the world and it would not be wrong to say that it is 'an enemy to humanity'. In this circumstance, the only option is preventing the occurrence of infection and preparing our healthcare system for the probable up-comings.

In that reference, it is extremely crucial to construct models that are computationally competent as well as realistic so that they can help policy makers, medical personals and also general public. Modeling the disease and providing future forecast of possible number of daily cases can assist the medical system in getting prepared for the new patients. The statistical prediction models are useful in forecasting as well as controlling the global epidemic threat.

In the present effort, we have employed Auto-Regressive Integrated Moving Average (ARIMA) model for predicting the incidence of 2019-nCov disease. As compared to other prediction models, for instance support vector machine (SVM) and wavelet neural network (WNN), ARIMA model is more capable in the prediction of natural adversities.[10] For our study, we have identified the best ARIMA model and then predicted the number the cases for the next 20 days. The main objective of the study is to find the best predictive model and apply it to forecast future incidence of COVID-19 cases in India.

## 2. METHODOLOGY

### 2.1 Dataset

Confirmed, recovered and death cases of COVID-19 infection are collected for India as well as countries with highest confirmed infection (US, Spain, Italy, France, Germany, China and Iran) and countries in South-East Asia region (India, Indonesia, Thailand, Bangladesh, Sri Lanka, Maldives, Nepal, Bhutan and Timor-Leste), as per World Health Organization region classification, from the official website of Johns Hopkins University (https://gisanddata.maps.arcgis.com/apps/opsdashboard/index.html) from 22 January 2020 to 13 April 2020.[4] This data is used to build predictive models.

### 2.2 Model Development

For forecasting a time series, ARIMA modeling is one of the best modeling techniques. ARIMA models are always represented with the help of some parameters and the model is expressed as ARIMA ($p$, $d$, $q$). Here, $p$ stands for the order of auto-regression, $d$ signifies the degree of trend difference while $q$ is the order of moving average. We have applied an ARIMA model to the time series data of confirmed COVID-19 cases in India. Autocorrelation function (ACF) graph and partial autocorrelation (PACF) graph is used to find the initial number of ARIMA models. These ARIMA models are then tested for variance in normality and stationary. Next, they are checked for accuracy by observing their MAPE, MAD and MSD values to determine the finest model to forecast. In addition, the best fit ARIMA model is compared with Linear Trend, Quadratic Trend, S-Curve Trend, Moving Average, Single Exponential as well as Double Exponential models using an output of measure of accuracy, *viz.* MAPE, MAD, MSD, so as to select the finest model to forecast. The finest model is the one which has the lowest value for all the measures. After fitting the model, its parameters are estimated followed by verification of the model. The built model is employed to forecast confirmed COVID-19 cases for the next 20 days, *i.e.* 14 April 2020 to 3 May 2020. The model for forecasting future confirmed COVID-19 cases is represented as,

$$ARIMA(p,d,f): X_t = \alpha_1 X_{t-1} + \alpha_2 X_{t-2} + \beta_1 Z_{t-1} + \beta_2 Z_{t-2} + Z_t \quad (1)$$

where
$$Z_t = X_t - X_{t-1} \quad (2)$$

Here, $X_t$ is the predicted number of confirmed COVID-19 cases at $t^{th}$ day, $\alpha_1$, $\alpha_2$, $\beta_1$ and $\beta_2$ are parameters whereas $Z_t$ is the residual term for $t^{th}$ day. The trend of forthcoming incidences can be estimated from the previous cases and a time series analysis is performed for this purpose. Time series forecasting refers to the employment of a model to forecast future data based on previously observed data.[11] In the present study, time series analysis is used to recognize the trends in confirmed COVID-19 cases in India over the period of 22 January 2020 to 13 April 2020 and to predict future cases from 14 April 2020 till 3 May 2020. The level of statistical significance is set at 0.05. A graph is plotted for actual confirmed cases and predicted confirmed cases with respect to time to verify the efficiency of the model. To get an idea of the recovery and death trends in India, a graph is plotted with respect to time.

A comparative study is also performed to examine the status of confirmed COVID-19 cases of India with respect to those of highly infected countries. A similar comparison is made with the countries of South-East Asia region as well. All the model developments, computations and comparisons have been performed using Minitab software (version 17).[12]

## 3. RESULT AND DISCUSSION

The present work encompasses development of a model to forecast COVID-19 incidences in the coming days. The results for measure of model accuracy for ARIMA, Linear Trend, Quadratic Linear, S-Curve Trend, Moving Average, Single Exponential as well as Double Exponential model are presented in Table 1. A look at the MAPE, MAD and MSD values suggests that ARIMA (2, 2, 2) model is the most accurate of all for forecasting future incidences as it possesses the least value for all the measures.

**Table 1** Measures of model accuracy

| Models | MAPE | MAD | MSD |
|---|---|---|---|
| ARIMA (2, 2, 2) | 4.1 | 58.3 | 25319.5 |
| Single exponential method | 9.7 | 98.3 | 58982.5 |
| Double exponential method | 9.1 | 52.2 | 18909.5 |
| Moving average (MA) | 10 | 141 | 103715 |
| S-Curve Trend Model | 66 | 1094 | 6798514 |
| Quadratic Trend Model | 13230 | 835 | 1085082 |
| Linear Trend Model | 14845 | 1300 | 3007510 |

Thus, parameters are estimated for the ARIMA (2, 2, 2) model which are displayed in Table 2. It is observed that AR (2) and MA (2) parameters have a *p*-value of 0.000, 0.167, 0.000 and 0.000 respectively, thus implying that the parameters are significant in the model.

**Table 2** Parameters estimates of the ARIMA model

| Type | Coeff | SE Coeff | t | p |
|---|---|---|---|---|
| AR (1) | 0.5363 | 0.1327 | 4.04 | 0.000 |
| AR (2) | -0.2048 | 0.1467 | -1.40 | 0.167 |
| MA (1) | 1.5656 | 0.0561 | 27.90 | 0.000 |
| MA (2) | -0.9486 | 0.0466 | -20.34 | 0.000 |

Figure 1 displays the residual plots for confirmed COVID-19 cases in India from 30 January 2020 to 13 April 2020. A slight deviation of residuals from the straight line can be observed from the plot. This indicates that the errors are somewhat near to normal with a few outliers. Therefore, the normality assumption is followed. The residual histogram backs up this assumption. The graph between residuals and the fitted values displays a little dispersion. This implies that the assumption of constant variance is also satisfied by the model. The non-correlation of residuals is clear from the plot of residuals *versus* the order of the data. The Ljung Box statistics further corroborate this fact (Table 3). It is transparent that the *p*-value of all the lags is larger than the significance level (*i.e.* 0.05) which means there is no violation of independent assumption. The suitability of the ARIMA (2, 2, 2) model is indicated by the non-significance of *p*-value and other statistics.

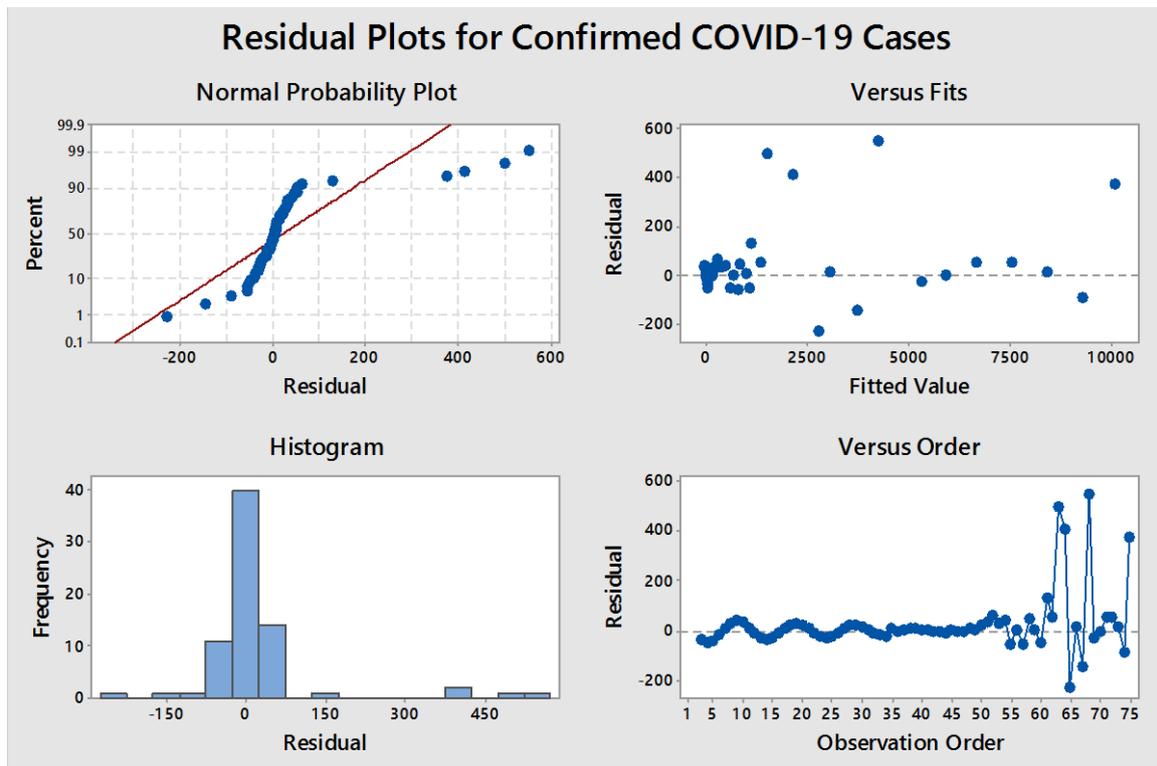

**Figure 1.** Residual plots for confirmed COVID-19 cases in India from 30 January 2020 to 13 April 2020.

**Table 3** Modified box-pierce (Ljung-Box) $\chi^2$ statistic

| Lag | $\chi^2$ | DF | *p*-Value |
|---|---|---|---|
| 12 | 20.5 | 8 | 0.009 |
| 24 | 21.1 | 20 | 0.391 |
| 36 | 21.6 | 32 | 0.918 |
| 48 | 23.6 | 44 | 0.995 |

Hence, the workable model obtained after the substitution of estimated parameters is represented as,

$$X_t = 0.5363 X_{t-1} - 0.2048 X_{t-2} + 1.5656 Z_{t-1} - 0.9486 Z_{t-2} + Z_t \qquad (3)$$

ARIMA (2, 2, 2) model (Eq. (3)) is used to forecast confirmed COVID-19 cases in India for the next 20 days, *i.e.* 14 April 2020 to 3 May 2020. The forecast for cases is presented in Table 4 with 95% confidence interval (CI). According to the forecast, the number of confirmed COVID-19 cases is expected to increase considerably in the coming 20 days. This increase is highly suspected to be associated with the people involved in a large social gathering which took place just before the lockdown was imposed. Many of them have been tested positive while a number of them are still untraceable. Thus, these people may cause transmissions and lead to higher number of infected figures. Another reason may be the negligence on the part of a few people who didn't follow the suggested 14 day isolation after returning from abroad. Further, there is still a possibility that the transmission might be occurring from asymptomatic cases with/without a travel history. It is also suspected that there are many asymptomatic cases which are still not tested. To some extent, social media is also contributing towards some cases owing to the fake information being spread through the platform. It is very important to control such communications as they result in people moving out of their places due to wrong informations. All these circumstances can end up in transmissions. Apart from that, until now, it hasn't been confirmed whether a recovered person can act as a carrier of the virus or not. Further, if it is possible, then for how long.

The lockdown period that began from 25 March 2020 has been extended up to 3 May 2020 considering the present cases according to the current update. It has been noted that some people have ignored the situation and warnings which resulted in a hasty increase in the number of infected cases. Hence, it is extremely crucial that people are made aware of the situation and the lockdown is strictly imposed in the whole country to prevent further transmission of the infection. If stringent measures are taken, it is believed that the number of new infected cases should begin declining in approximately 20 days. Looking at the prevention approach employed by China, that is, severe control and quarantine, it can be expected that India will also recover soon because of its similar preventive measures.

**Table 4** Figures for forecasted confirmed COVID-19 cases and their lower and upper limits for 20 days (14 April 2020 to 3 May 2020) with 95% CI

| Date | Forecast | Lower limit | Upper limit |
|---|---|---|---|
| 14-Apr-20 | 11307.5 | 11070.1 | 11545.0 |
| 15-Apr-20 | 12208.3 | 11877.3 | 12539.2 |
| 16-Apr-20 | 13214.4 | 12787.8 | 13640.9 |
| 17-Apr-20 | 14267.5 | 13694.8 | 14840.2 |
| 18-Apr-20 | 15324.2 | 14546.7 | 16101.7 |
| 19-Apr-20 | 16373.3 | 15346.5 | 17400.1 |
| 20-Apr-20 | 17417.5 | 16109.7 | 18725.3 |

| Date | | | |
|---|---|---|---|
| 21-Apr-20 | 18460.7 | 16846.7 | 20074.7 |
| 22-Apr-20 | 19504.3 | 17561.7 | 21447.0 |
| 23-Apr-20 | 20548.4 | 18256.1 | 22840.8 |
| 24-Apr-20 | 21592.6 | 18930.8 | 24254.4 |
| 25-Apr-20 | 22636.8 | 19586.9 | 25686.8 |
| 26-Apr-20 | 23681.0 | 20225.1 | 27136.9 |
| 27-Apr-20 | 24725.2 | 20846.4 | 28603.9 |
| 28-Apr-20 | 25769.3 | 21451.4 | 30087.2 |
| 29-Apr-20 | 26813.5 | 22040.8 | 31586.1 |
| 30-Apr-20 | 27857.6 | 22615.1 | 33100.2 |
| 01-May-20 | 28901.8 | 23174.8 | 34628.8 |
| 02-May-20 | 29946.0 | 23720.3 | 36171.6 |
| 03-May-20 | 30990.1 | 24252.0 | 37728.2 |

Time series analysis presents the meaningful statistics for confirmed COVID-19 data. Figure 2 presents the time series graph of the active infected COVID-19 cases from 30 January 2020 to 3 May 2020. It is clear from the plot that the time series is not stationary. An increasing trend is displayed by the time series suggesting a high rise in COVID-19 cases.

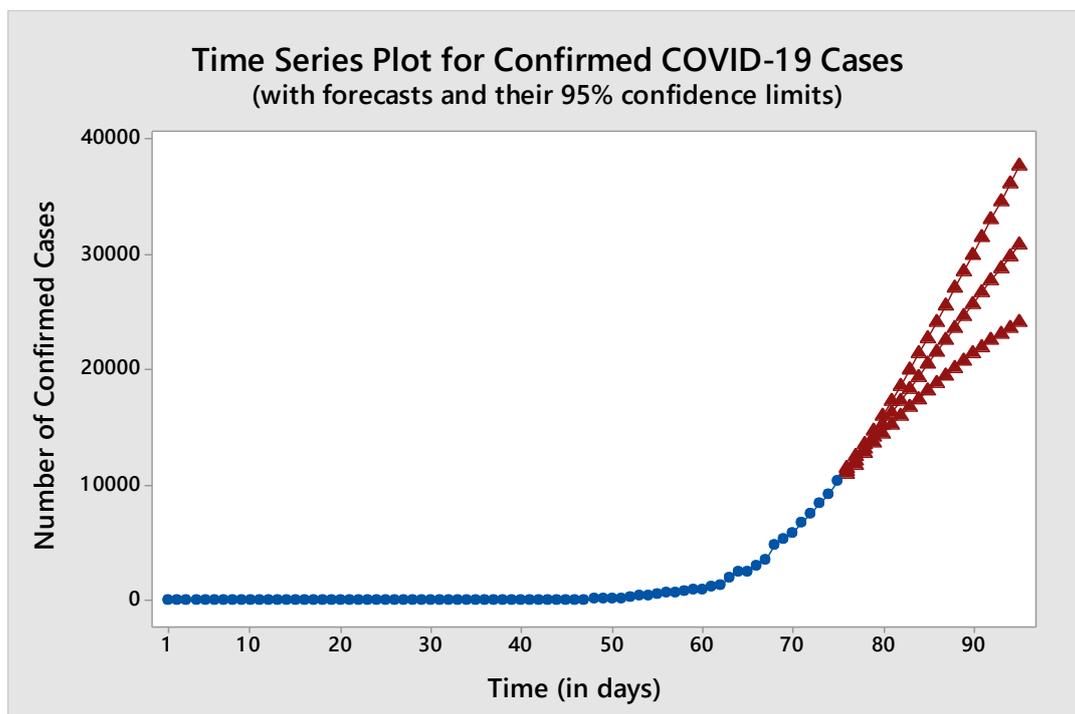

**Figure 2.** Times series plot for confirmed COVID-19 infections in India from 30 January 2020 to 3 May 2020 (Blue line represents actual confirmed cases[4] and red lines represent case forecasts).

For comparing the actual and forecasted confirmed COVID-19 cases, a time series graph is plotted starting from 30 January 2020 till 13 April 2020. The plot is represented by Figure 3. The similarity of forecasted data with actual data is clear from these plots. This comparison reveals the precision of the model in forecasting.

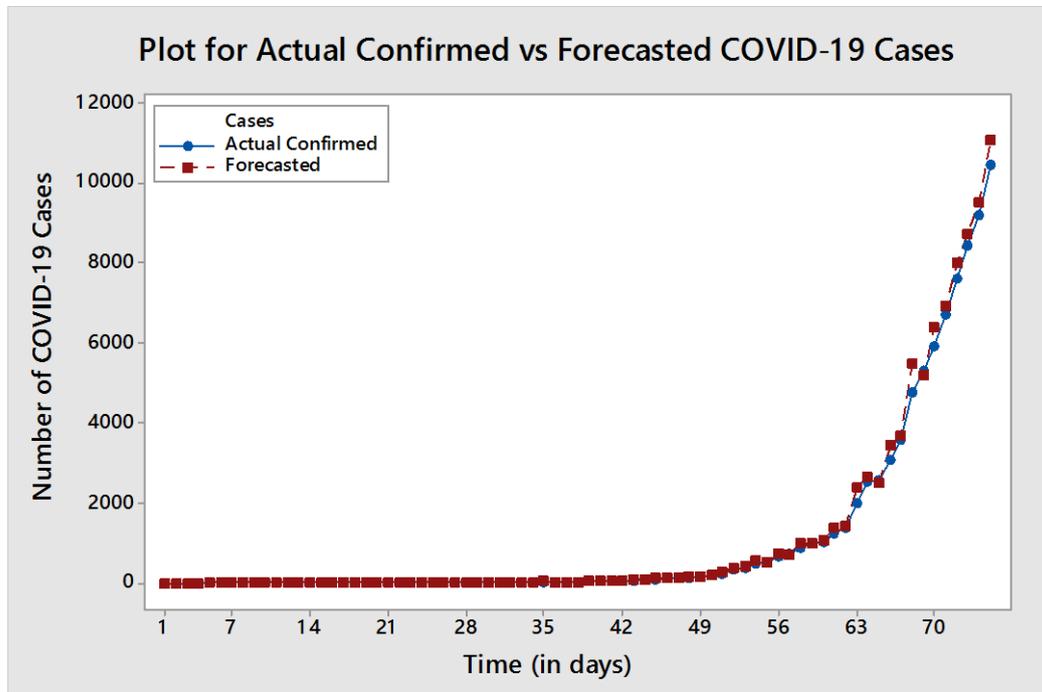

**Figure 3.** A comparative times series plot for actual confirmed[4] and forecasted COVID-19 cases from 30 January 2020 to 13 April 2020.

Trend for the number of recovery and death cases with respect to time due to COVID-19 infections in India depicted in Figure 4. It is observed that the number of recoveries as well as deaths increase with time, however the rate of recovery is higher than the death rate. Thus, a low mortality rate could be expected from the disease.

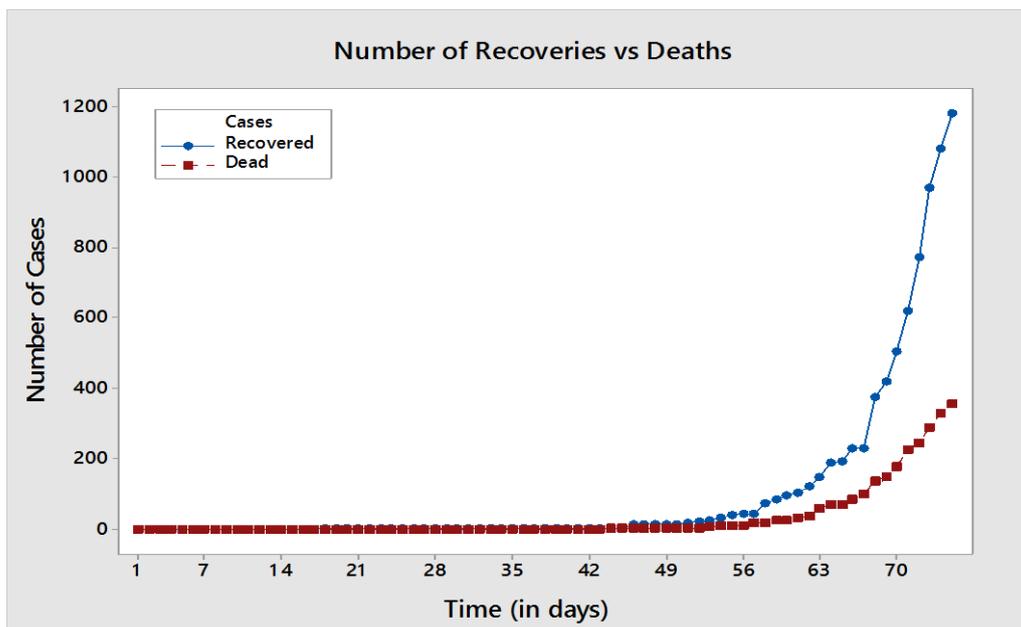

**Figure 4.** A comparative trend for the number of recoveries and deaths due to COVID-19 infections in India from 30 January 2020 to 13 April 2020[4].

Figure 5 shows a comparative study of confirmed COVID-19 infection cases of India with respect to those of highly infected countries. According to the plot, US is the most infected while India the least infected of the selected countries, *viz.* US, Spain, Italy, France, Germany, China and Iran. It is very obvious as India was the last amongst these countries to get infected. However, the plot also reflects that China has been able to control the pandemic and is now presenting very few new cases. Thus, it follows that if strict prevention measures such as quarantine and sanitization are continued for some days, the situation could be controlled in the coming days. In the remaining countries, infected cases are growing exponentially and severe spread of infection is seen.

A similar comparison is performed for the countries of South-East Asia region as well as shown in Figure 6. A look at the Figure 6 suggests India to be the most infected amongst the South-East Asian countries followed by Indonesia and Thailand. All the three countries are presenting continuous rise in confirmed COVID-19 infections. The remaining countries of the region have a very low infection rate, lowest being in Timor-Leste.

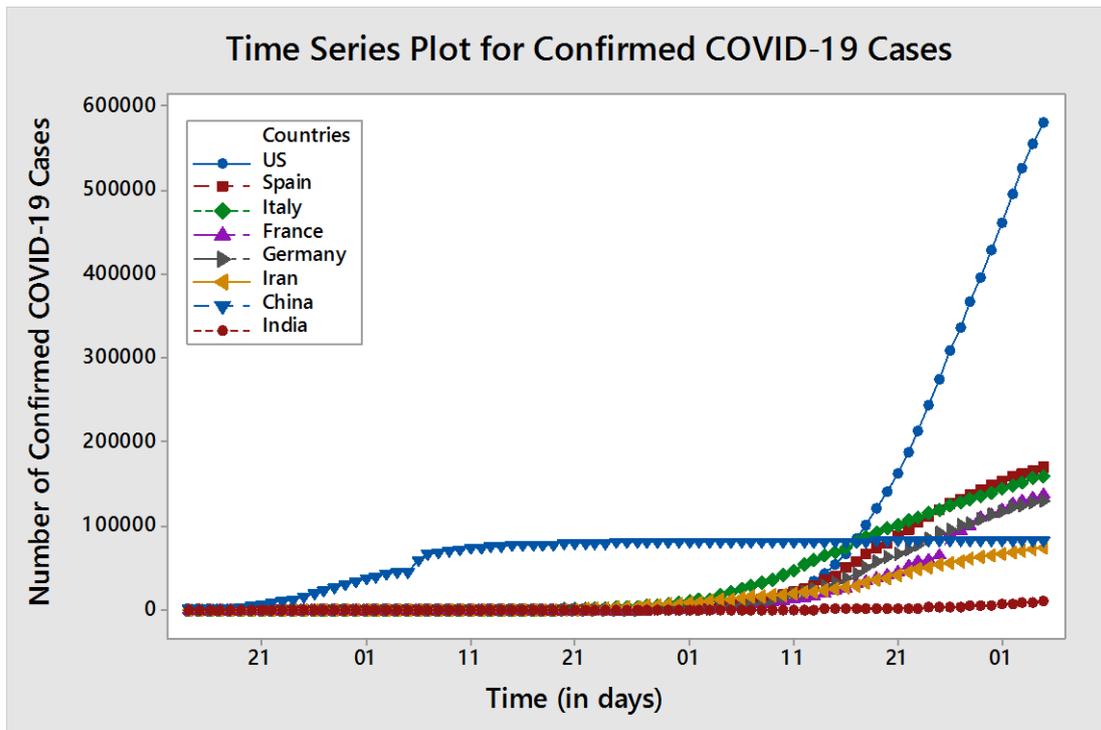

**Figure 5.** A comparative plot of confirmed COVID-19 cases in China, Italy, US, Spain, Germany, Iran and India from 22 January 2020 to 13 April 2020[4].

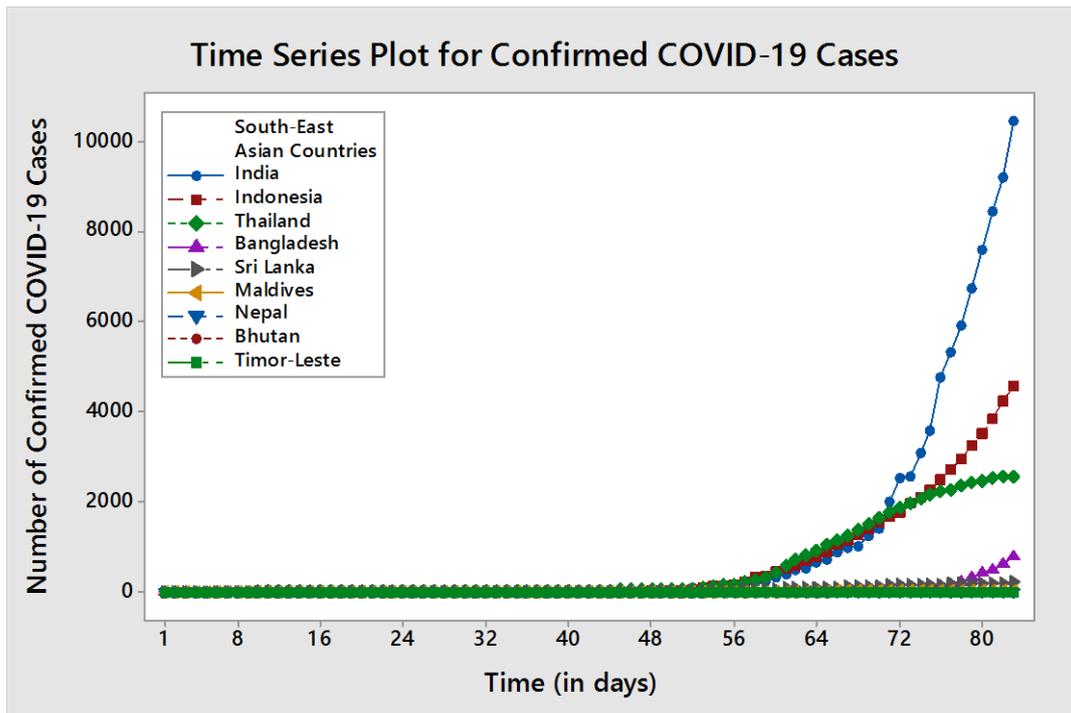

**Figure 6.** A comparative plot of confirmed COVID-19 infection cases for the countries in South-East Asia region from 22 January 2020 to 13 April 2020[4].

It is clear that measures like quarantine and sanitization can decrease human exposure and control this pandemic. Thus, these measures should be stringently imposed in India and strict actions must be taken against those people who violate the rules and don't consider the severity of the situation. Although a large amount of data helps in providing a more exhaustive prediction and explanation, in the present circumstance, these models could be valuable in anticipating future cases of infection if the pattern of virus spread didn't change abnormally. It is obvious that this virus is new and has the capability to be transmitted intensely. Hence, it may have an influence on the predictions, however as per our knowledge, in the present situation this model is the finest.

## 4. CONCLUSION

The novel coronavirus disease (COVID-19) has been declared as pandemic by WHO and is currently a major global threat. In order to support the prevention of the disease and aid in the healthcare service preparation, we have conducted this study to examine the finest model for the prediction of confirmed COVID-19 infection cases and to employ that model for forecasting future COVID-19 infection cases in India. As per the model forecast, the confirmed cases are expected to greatly rise in the coming days. The time series analysis shows an exponential enhancement in the infected cases. However, it is also anticipated that the efforts such as lockdown may affect this prediction and cases may start to decline after a month approximately. A comparative study with some of the highly infected countries and countries in south-east Asia region indicates that India can still control the situation if the prevention measures such as quarantine and city sanitization are strictly followed. The prediction models will help the government and medical workforce to be prepared for the upcoming situations and have more readiness in healthcare systems.

## ACKNOWLEDGEMENTS

Both the corresponding authors are thankful to Presidency University, Bengaluru and Manipal University Jaipur, Jaipur for providing research facility.

## AUTHOR CONTRIBUTIONS

H.T. and T.C. conceptualized the project. H.T. designed the study, performed the computations and investigations, contributed to data analysis and wrote the manuscript. P.R. provided the resources. T.C. and V.S. supervised the study and reviewed the manuscript.

## COMPETING INTERESTS

The authors declare no competing interests.

## FUNDING

This research did not receive any specific grant from funding agencies in the public, commercial, or not-for-profit sectors.

## REFERENCES

1. World Health Organization, Coronavirus disease (COVID-19) outbreak. https://www.who.int/emergencies/diseases/novel-coronavirus-2019 (accessed on April 14, 2020).
2. Zhu, N.; Zhang, D.; Wang, W.; Li, X.; Yang, B.; Song, J.; Zhao, X.; Huang, B.; Shi, W.; Lu, R.; Niu, P.; Zhan, F.; Ma, X.; Wang, D.; Xu, W.; Wu, G.; Gao, G. F.; Tan, W. A novel coronavirus from patients with pneumonia in China, 2019. *N. Engl. J. Med.* **2020**, 382, 727.
3. Paules, C. I.; Marston, H. D.; Fauci, A. S. Coronavirus infections—more than just the common cold, *JAMA* **2020**, 323, 707.
4. Johns Hopkins University Center for Systems Science and Engineering, Coronavirus (COVID-19) Cases. https://github.com/CSSEGISandData/COVIDQ3 (accessed on April 14, 2020).
5. Wikipedia, 2019-20 coronavirus outbreak. https://en.wikipedia.org/wiki/2019-20_coronavirus_outbreak (accessed on April 14, 2020).
6. N. C. P. E. R. E. Team. The epidemiological characteristics of an outbreak of 2019 novel coronavirus diseases (COVID-19) in China, *China CDC Weekly* **2020**, 41, 145.
7. Huang, C.; Wang, Y.; Li, X.; Ren, L.; Zhao, J.; Hu, Y.; Zhang, L.; Fan, G.; Xu, J.; Gu, X.; Cheng, Z.; Yu, T.; Xia, J.; Wei, Y.; Wu, W.; Xie, X.; Yin, W.; Li, H.; Liu, M.; Xiao, Y.; Gao, H.; Guo, L.; Xie, J.; Wang, G.; Jiang, R.; Gao, Z.; Jin, Q.; Wang, J.; Cao, B. Clinical features of patients infected with 2019 novel coronavirus in Wuhan, China, *The Lancet* **2020**, 395, 497.
8. Niehus, R.; De Salazar, P. M.; Taylor, A.; Lipsitch, M. Quantifying bias of COVID-19 prevalence and severity estimates in Wuhan, China that depend on reported cases in international travellers, *medRxiv*. **2020**.

9. Unnithan, P. S. G. Kerela confirmed first novel coronavirus case in India, *India Today*. https://www.indiatoday.in/india/story/kerala-reports-first-confirmed-novel-coronavirus-case-in-india-1641593-2020-01-30 (2020) (accessed on April 14, 2020).
10. Zhang, Y.; Yang, H.; Cui, H.; Chen, Q. Comparison of the Ability of ARIMA, WNN and SVM Models for Drought Forecasting in the Sanjiang Plain, China. *Nat. Resour. Res.* **2019**, 29, 1447.
11. Imdadullah, M. *Time series analysis. Basic statistics and data analysis* **2014**.
12. Minitab, Inc., State College, PA. Minitab 17 Statistical Software (**2010**).